\documentclass[noshowpacs,amsmath,
twocolumn,
superscriptaddress,
8pt,
aps,prb]{revtex4-2}

\usepackage{setspace}
\usepackage{amsmath}
\usepackage{booktabs}
\usepackage{multirow}
\usepackage{graphicx}

\usepackage{verbatim}
\usepackage{amsfonts}
\usepackage{amssymb}
\usepackage{epstopdf}
\usepackage{dcolumn}
\usepackage{xcolor}
\usepackage{verbatim}
\setcitestyle{super}
\DeclareGraphicsExtensions{.pdf,.eps,.png,.jpg,.mps}
\usepackage{textcomp}
\usepackage{upgreek}
\usepackage{siunitx}

\begin{document} 


\title{Large-scale cluster quantum microcombs}

\author
{Ze Wang,$^{1,\ast}$ Kangkang Li,$^{1,\ast,\dagger}$ Yue Wang,$^{1,\ast}$ Xin Zhou,$^{2}$ Yinke Cheng,$^{1,2}$ Boxuan Jing,$^{1}$ Fengxiao Sun,$^{1}$ \\
Jincheng Li,$^{2}$ Zhilin Li,$^{1}$ Bingyan Wu,$^{2}$ Qihuang Gong,$^{1,3,4,5}$ Qiongyi He,$^{1,3,4,5\dagger}$ Bei-Bei Li,$^{2,\dagger}$ and Qi-Fan Yang$^{1,3,4,5\dagger}$\\
$^{1}$State Key Laboratory for Artificial Microstructure and Mesoscopic Physics and Frontiers Science Center for Nano-optoelectronics School of Physics, Peking University, Beijing, 100871, China\\
$^2$Beijing National Laboratory for Condensed Matter Physics, Institute of Physics, Chinese Academy of Sciences, Beijing, 100190, China\\
$^3$Collaborative Innovation Center of Extreme Optics, Shanxi University, Taiyuan, 030006, China\\
$^4$Peking University Yangtze Delta Institute of Optoelectronics, Nantong, Jiangsu, 226010, China\\
$^5$Hefei National Laboratory, Hefei, 230088, China\\
$^{\ast}$These authors contributed equally to this work.\\
$^{\dagger}$Corresponding author: kangkangli@pku.edu.cn; qiongyihe@pku.edu.cn; libeibei@iphy.ac.cn; leonardoyoung@pku.edu.cn.
}

\maketitle


\noindent{\bf An optical frequency comb comprises a cluster of equally spaced, phase-locked spectral lines. Replacing these classical components with correlated quantum light gives rise to cluster quantum frequency combs, providing abundant quantum resources for measurement-based quantum computation and multi-user quantum networks. We propose and generate cluster quantum microcombs within an on-chip optical microresonator driven by multi-frequency lasers. Through resonantly enhanced four-wave mixing processes,  continuous-variable cluster states with 60 qumodes are deterministically created. The graph structures can be programmed into one- and two-dimensional lattices by adjusting the configurations of the pump lines, which are confirmed inseparable based on the measured covariance matrices. Our work demonstrates the largest-scale cluster states with unprecedented raw squeezing levels from a photonic chip, offering a compact and scalable platform for computational and communicational tasks with quantum advantages.
}

\medskip

Quantum entanglement, as conceptualized by Einstein, Podolsky, Rosen (EPR) \cite{einstein1935can}, and Schr$\mathrm{\Ddot{o}}$dinger \cite{schrodinger1935discussion}, has spurred numerous quantum technologies that promise superior speed and security over classical counterparts.  One key metric to these applications is the number of quantum entities involved. When these entities are entangled in specific graph structures called cluster states, it enables the implementation of novel quantum protocols, including measurement-based quantum computing \cite{raussendorf2003measurement,menicucci2006universal} and unconditional quantum teleportation \cite{furusawa1998unconditional}. Optical photons have been demonstrated to be versatile for quantum computing and long-distance quantum information transmission. However, the probabilistic nature of single-photon sources has limited the construction of large-scale cluster states, with the current maximum being 12 photons \cite{zhong201812} or 18 photonic qubits \cite{PhysRevLett.120.260502}.

Instead of qubits, EPR's original concept of quantum entanglement was encoded in continuous variables (CVs), such as the positions and momenta of particles \cite{einstein1935can}. This form of quantum entity, known as quantum modes (qumodes), can be deterministically entangled via Gaussian operations \cite{braunstein2005quantum,weedbrook2012gaussian}. In quantum optics, these CVs are typically chosen as quadrature components, defined as \(\hat{x} = (\hat{a} + \hat{a}^\dagger) / \sqrt{2}\) and \(\hat{p} = -i(\hat{a} - \hat{a}^\dagger) / \sqrt{2}\), with \(\hat{a}\) (\(\hat{a}^\dagger\)) being the annihilation (creation) operator of the quantized electromagnetic field. CV quantum states can be represented using the correlations between these quadratures in the form of covariance matrices.

The building blocks for photonic CV cluster states are squeezed vacuums generated by optical parametric amplifiers or oscillators. Multiplexing them in spectral \cite{pysher2011parallel,roslund2014wavelength,chen2014experimental,cai2017multimode}, temporal \cite{yokoyama2013ultra,asavanant2019generation,larsen2019deterministic,madsen2022quantum}, path \cite{su2013gate,arrazola2021quantum}, and spatial \cite{armstrong2012programmable,zhang2020reconfigurable} domains has resulted in large-scale cluster states containing up to 1 million qumodes. However, aiming for compactness and scalability, realizing large-scale CV cluster states on photonic chips encounters two significant challenges. First, chip-based squeezers, including waveguide parametric amplifiers \cite{kanter2002squeezing,lenzini2018integrated,kashiwazaki2020continuous} and nonlinear optical microresonators \cite{otterpohl2019squeezed,zhao2020near,vaidya2020broadband,zhang2021squeezed,yang2021squeezed,jahanbozorgi2023generation,park2024single,9890809}, underperform their tabletop counterparts in raw squeezing -- a key performance metric that measures the reduced quadrature noise relative to the vacuum noise as received by photodetectors. Second, the maximum number of identical squeezers on a single photonic chip is currently limited by fabrication yield. To date, the largest cluster states generated on a photonic chip are capped at 8 qumodes \cite{arrazola2021quantum}. 

\begin{figure*}[hbtp]
\centering
\includegraphics[width=1.0\linewidth]{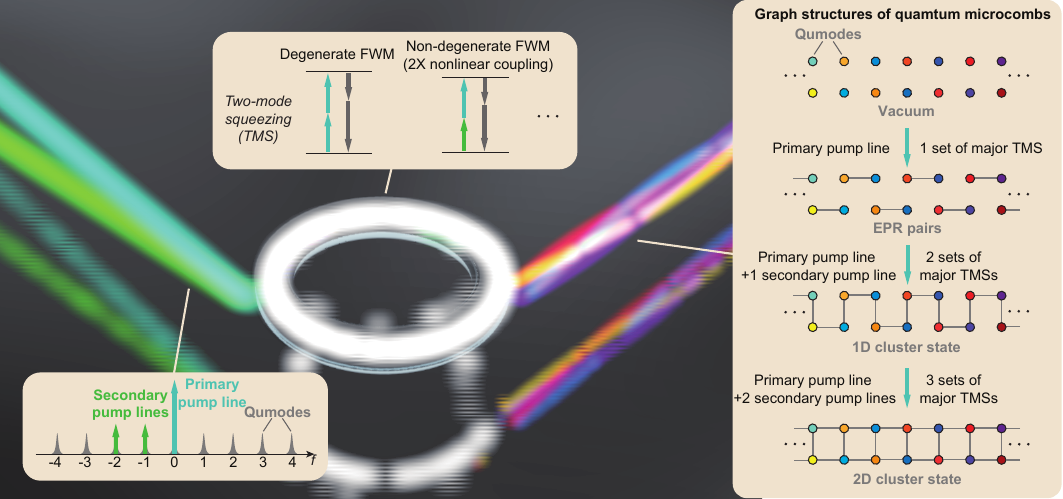}
\caption{{\bf Schematic illustration of cluster quantum microcombs.} The microresonator hosts many spectral qumodes, and several qumodes are simultaneously pumped by equally spaced continuous-wave lasers. Quantum microcombs with different entanglement structures are generated by two-mode squeezing (TMS), induced by either degenerate or non-degenerate four-wave mixing (FWM) processes.}
\label{figure1}
\end{figure*}

In this work, we demonstrate large-scale cluster states in a standalone on-chip microresonator squeezer providing state-of-the-art raw squeezing ($>3$ dB). The cluster states are quantum analogues of microresonator frequency combs (microcombs) due to the theoretically infinite, equally spaced spectral qumodes involved \cite{pysher2011parallel,roslund2014wavelength,chen2014experimental,cai2017multimode,yang2021squeezed}. Provided the high-quality ($Q$) factors and small mode volumes of the microresonator, Kerr-induced four-wave mixing (FWM) processes can efficiently establish quantum correlations among the qumodes. We devise FWM processes to construct 60-mode cluster quantum microcombs into one-dimensional (1D) and two-dimensional (2D) lattices, and the scale can be further expanded using dispersion-engineered devices.

\begin{figure*}[hbtp]
\centering
\includegraphics[width=1.0\linewidth]{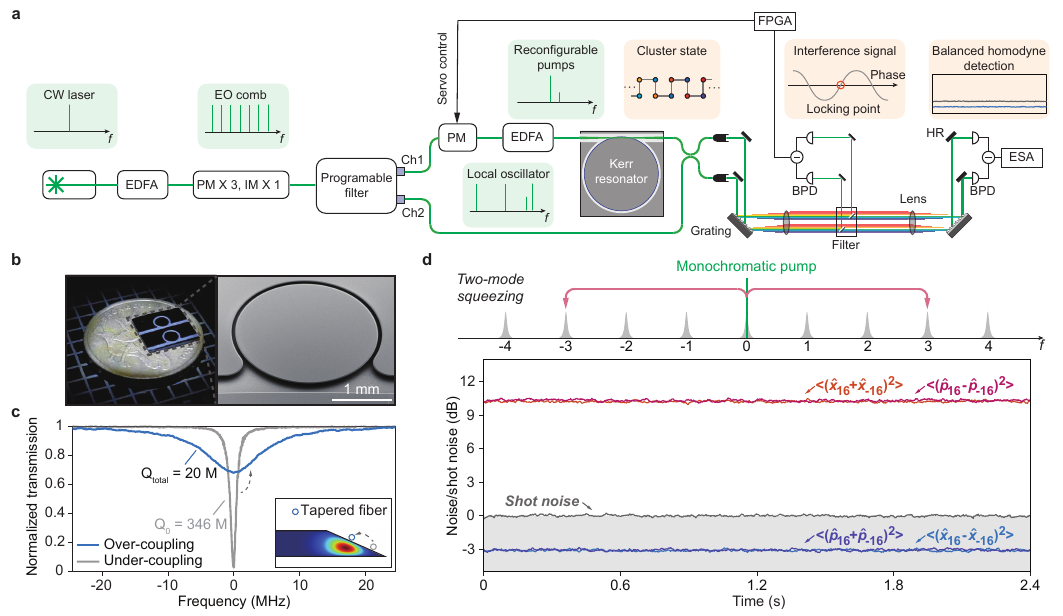}
\caption{{\bf Experimental setup and performance benchmark.} {\bf a,} Experimental setup. CW laser: Continuous-wave laser; EO comb: Electro-optic comb; EDFA: Erbium-doped fiber amplifier; PM: Phase modulator; IM: Intensity modulator; BPD: Balanced photodetector; ESA: Electrical spectrum analyzer; FPGA: Field-programmable gate array.
{\bf b,} Photo (left) and scanning-electron-microscopy (right) images of the microresonators. {\bf c,} Transmission spectrum of the microresonator. Inset: Cross-sectional profiles of the microresonator and the optical mode, with the tapered fiber coupler indicated. {\bf d,} Upper panel: Pumping scheme and FWM processes. Lower panel: Electrical spectra showing quadrature noise variance relative to shot noise (gray) for the qumode pair (-16, 16).}
\label{figure2}
\end{figure*}

\medskip
{\noindent\bf Results}

{\noindent\bf Principle of cluster quantum microcombs}

\noindent The protocol for generating cluster quantum microcombs is described in Fig. \ref{figure1}. The microresonator hosts a series of longitudinal modes separated by the free spectral range (FSR). Equally spaced pump lines are coupled to the microresonator, with their spacing ($f_r$) closely matched to the FSR to ensure resonant excitation. Qumodes are uniformly defined in the spectral domain with identical spacing $f_r$. The large FSR of microresonators ($>$10 GHz) provides line-by-line control of pump lines and evaluation of individual qumodes.

Within the microresonator, multiple FWM processes can occur; however, our focus is on two-mode squeezing (TMS), which induces quantum correlations between the quadratures of qumode pairs. The TMS interactions are described by the Hamiltonian
\begin{equation}
\begin{split}
    \hat{H} =& -\sum_{n}\sum_{k} g A_n^2 \hat{a}_k^{\dagger} \hat{a}_{-k+2n}^{\dagger} \\
    &- \sum_{n<m}\sum_{k} 2 g A_n A_m \hat{a}_k^{\dagger} \hat{a}_{-k+n+m}^{\dagger} + \mathrm{h.c.},
\end{split}  
\end{equation}
where $A_n$ denotes the amplitude of the pump line at the $n_\mathrm{th}$ qumode, $g$ is related to Kerr nonlinearity, and $\hat{a}_k$ ($\hat{a}_k^{\dagger}$) represent the annihilation (creation) operators for the $k_{\mathrm{th}}$ qumode. The factor of 2 in the second term accounts for non-degenerate FWM. The resulting quantum states depend on the power and arrangement of the pump lines. To graphically represent these states, we connect correlated qumodes using edges, with the edge weights indicating the strength of the quantum correlations. For example, utilizing only one pump line generates a single set of TMS, producing correlated EPR pairs known as squeezed quantum microcombs \cite{yang2021squeezed}.

Cluster states with equal-weight edges have been widely proposed in various quantum applications\cite{Menicucci2006Uni,van2011implementing}. To construct such states, it is essential to ensure that the strongest sets of TMSs (referred to as major TMSs) possess comparable coupling coefficients (i.e., $g A_n^2$ for degenerate FWM and $2 g A_n A_m$ for non-degenerated FWM). Our approach to generating multiple sets of major TMSs involves introducing a primary pump line alongside several weaker secondary pump lines. Intuitively, by tuning the power of the secondary pump lines to approximately $25\%$ of that of the primary pump line, the secondary pumps can create TMS through non-degenerate FWM with the primary pump line. These results in coupling coefficients are comparable to those induced by degenerate FWM from the primary pump line alone, thereby linking the qumodes to form a 1D cluster state with approximately equal-weighted connecting edges. Similarly, employing two secondary pump lines enables three sets of major TMSs, allowing the formation of 2D lattice geometries. Additional secondary pump lines can be utilized to realize customized graphs with equal-weight edges; however, the power must be adjusted accordingly, as a single set of major TMS can be contributed by multiple FWM processes. Detailed configurations of the pump lines to achieve equal weights and minimize other parasitic nonlinear processes are discussed in Supplementary Information.

\begin{figure*}
\centering
\includegraphics[width=1.0\linewidth]{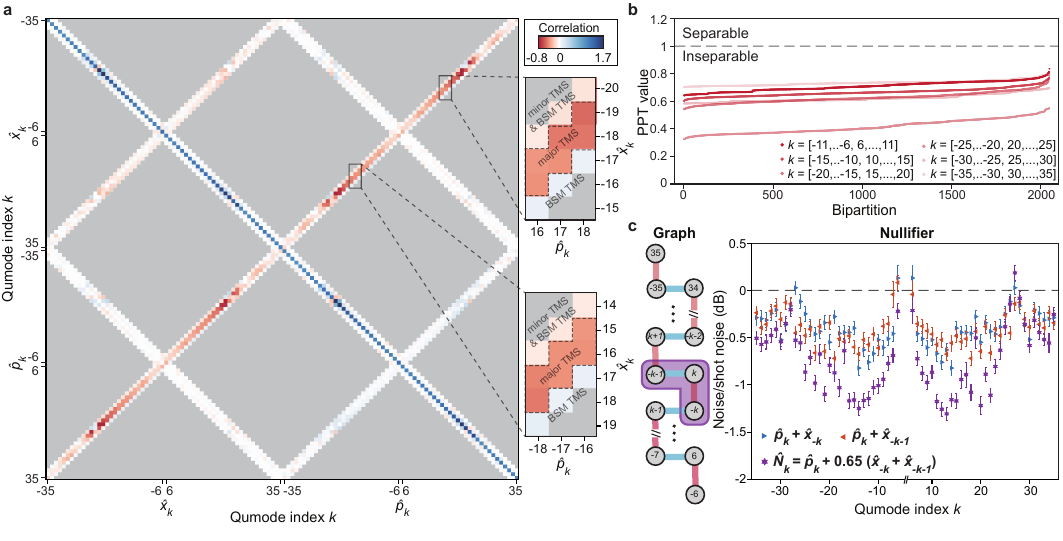}
\caption{{\bf 1D cluster quantum microcombs.} 
{\bf a,} Experimentally measured 60-mode covariance matrix. Measurements in the gray shades are not taken. Quantum correlations induced by the major TMSs are indicated between the dashed lines in the zoom-in views. {\bf b,} PPT inseparability criteria for 2047 bipartitions for 6 sets of 12 qumodes. All bipartitions with PPT values below 1 indicate complete inseparability of the state. {\bf c,} Graph structures (left panel) and the measured squeezing of nullifiers \(\hat{N}_k=\hat{p}_k+0.65 (\hat{x}_{-k}+\hat{x}_{-k-1})\) (right panel). The colors of the edges indicate different sets of major TMSs, and the purple shading indicates the nullified. Quadrature noise variances for $\hat{p}_k+\hat{x}_{-k}$ and $\hat{p}_k+\hat{x}_{-k-1}$ are also plotted for comparison.}
\label{figure3}
\end{figure*}

\begin{figure*}
\centering
\includegraphics[width=1.0\linewidth]{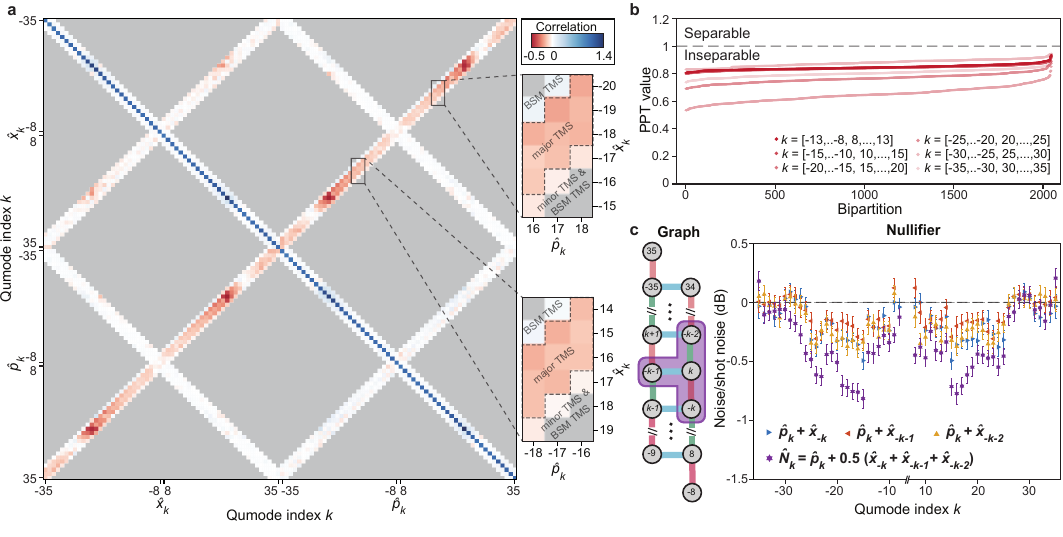}
\caption{{\bf 2D cluster quantum microcombs.} 
{\bf a,} Experimentally measured 56-mode covariance matrix. Measurements in the gray shades are not taken. Quantum correlations induced by the major TMSs are indicated between the dashed lines in the zoom-in views. {\bf b,} PPT inseparability criteria for 2047 bipartitions for 6 sets of 12 qumodes. All bipartitions with PPT values below 1 indicate complete inseparability of the state. {\bf c,} Graph structures (left panel) and the measured squeezing of nullifiers \(\hat{N}_k=\hat{p}_k+0.5(\hat{x}_{-k}+\hat{x}_{-k-1}+\hat{x}_{-k-2})\) (right panel). The colors of the edges indicate different sets of major TMSs, and the purple shading indicates the nullifier. Quadrature noise variances for $\hat{p}_k+\hat{x}_{-k}$, $\hat{p}_k+\hat{x}_{-k-1}$, and $\hat{p}_k+\hat{x}_{-k-2}$ are also plotted for comparison.}
\label{figure4}
\end{figure*}

\medskip

{\noindent\bf Experimental setup}

\noindent The experimental setup, illustrated in Fig. \ref{figure2}a, employs a 70-line electro-optic comb generated from a continuous-wave laser via sequential phase and amplitude modulation \cite{ishizawa2011generation,metcalf2013high}. A programmable multi-port filter selectively controls the phases and amplitudes of specific comb lines. One port provides multi-frequency pump lines directed into the microresonator. The second port functions as a local oscillator (LO) and shares 10\% of the power of the primary pump line, which is combined with the microresonator emission through a 50/50 beam splitter. The residual pump lines are filtered out prior to balanced homodyne detection. The electrical signals at the detector are expressed as \(\sum_k e^{-i\psi}E_k \hat{a}_k + \text{h.c.}\), where \(E_k\) represents the classical coherent fields in the \(k_\mathrm{th}\) mode of the LO and \(\psi\) is the relative phase between the two arms before the beam splitter. To stabilize $\psi$, the interference signal of the primary pump lines is used as the error signal for feedback controlling the phase modulator inserted before the microresonator (Fig. \ref{figure2}a). We can then stably access different quadratures by setting the LO using the programmable filter. Each quadrature noise variance is analyzed using an electrical spectral analyzer at 0.5 MHz frequency, 100 kHz resolution bandwidth, and 10 Hz video bandwidth. A measurement without input from the quantum microcomb is performed as the corresponding shot noise of the LO, based on which the raw squeezing is calculated. Further details of the experimental setup are provided in Fig.~S10 of Supplementary Information. 

The microresonator is a silica disk fabricated on a silicon chip \cite{lee2012chemically} with an FSR of approximately 25 GHz (Fig. \ref{figure2}b). Given an intrinsic \(Q\) exceeding 300 million, the position of the fiber taper is adjusted to achieve an over-coupling condition with a total \(Q\) of 20 million, resulting in an extraction efficiency of 94\% (Fig. \ref{figure2}c). Considering additional parasitic losses during filtering and detection, the overall detection efficiency of our system is estimated to be near 67.5\%. To benchmark the performance of our setup, we conduct two-mode squeezing tests using only the primary pump line. Figure \ref{figure2}d presents the electrical spectra showing the quadrature noise variance for qumodes (-16, 16), normalized to the shot noise. The maximally squeezed quadratures, \(\hat{x}_{16} + \hat{x}_{-16}\) and \(\hat{p}_{16} - \hat{p}_{-16}\), exhibit raw squeezing (corrected for the photodetector’s electronic noise) of \(3.08 \pm 0.07\) dB and \(3.04 \pm 0.10\) dB, respectively. Conversely, the maximally anti-squeezed quadratures, \(\hat{x}_{16} - \hat{x}_{-16}\) and \(\hat{p}_{16} + \hat{p}_{-16}\), reach raw anti-squeezing levels of \(10.29 \pm 0.07\) dB and \(10.15 \pm 0.07\) dB, respectively. A comprehensive characterization of squeezing across 31 EPR pairs is provided in Fig.~S13 of Supplementary Information.

\medskip

\noindent\textbf{1D cluster quantum microcombs}

\noindent We generate 1D cluster quantum microcombs using a primary pump line and a secondary pump line. The secondary pump is applied to mode $-1$ with a power set to $20\%$ instead of $25\%$ of the primary pump power to suppress the generation of undesired sidebands near the pump lines (see Supplementary Information). The cluster state is characterized by measuring its covariance matrix defined on the $\hat{x}$ and $\hat{p}$ quadratures of different qumodes. The covariance matrix for 60 qumodes (120 quadratures), with indices ranging from $\pm6$ to $\pm35$, is presented in Fig.~\ref{figure3}a. Due to its large size, we selectively measure only the elements expected to exhibit correlations (i.e., non-zero), which are validated through zero-checks (see Fig.~S14 in Supplementary Information). The phases of the measurement basis are realigned to enhance the visualization of correlations (see Supplementary Information). Negative correlations induced by two sets of major TMSs are observed along the off-diagonal directions for $(\hat{p}_k, \hat{x}_{-k})$ and $(\hat{p}_k, \hat{x}_{-k-1})$. Additional quantum correlations arise from two sources: minor TMS generated solely by the secondary pump on $(\hat{p}_k, \hat{x}_{-k-2})$ and Bragg scattering, as described by the Hamiltonian $-2g A_0 A_{-1}^* \hat{a}_{-k+1}^{\dagger} \hat{a}_{-k} + \mathrm{h.c.}$. Although Bragg scattering does not directly induce quantum correlations, it cooperates with major TMS to generate correlations between $(\hat{p}_k, \hat{x}_{-k+1})$ and $(\hat{p}_k, \hat{x}_{-k-2})$. Notably, quantum correlations induced by both minor TMS and Bragg-scattering-mediated (BSM) TMS are significantly weaker compared to those induced by major TMSs. Therefore, the graph structures are primarily determined by the major TMSs.

We employ the Positive Partial Transpose (PPT) criterion \cite{Peres1996Separability,Horodecki1996Separability,Simon2000Peres} to evaluate the inseparability of the cluster states (Fig.~\ref{figure3}b). The PPT values are calculated for all possible bipartitions of the states. For our 60-mode system, the total number of bipartitions is $2^{59} - 1$, which is computationally challenging. Therefore, the cluster states are partitioned into six subsets, each comprising twelve qumodes and 2047 bipartitions. The PPT values calculated from the covariance matrix are below unity for these bipartitions, indicating inseparability. Given the overlap between these subsets, the entire cluster state is deemed inseparable.

The graph structure inferred from the major TMSs is a 1D lattice (Fig.~\ref{figure3}d), with edge weights determined by nullifiers -- a set of orthogonal operators for which the cluster state is an eigenstate with zero eigenvalues. They have the standard form $\hat{p}_k - \sum_{n \neq k} \mathbf{V}_{nk} \hat{x}_k$, where $\mathbf{V}_{nk}$ are the weight of the edge connecting qumodes $n$ and $k$. Although nullifiers are defined to feature zero variance (infinite squeezing), their approximations can be found in practical systems with finite squeezing. $\hat{N}_k=\hat{p}_k+0.65(\hat{x}_{-k}+\hat{x}_{-k-1})$ are found as good approximations of nullifiers due to their decent squeezing inferred from the covariance matrix (see Fig.~S16 in Supplementary Information). As an experimental proof, we directly measure the quadrature noise variances of 60 nullifiers. The nullifiers exhibit squeezing of up to $1.31 \pm 0.07$ dB below the shot noise level, while the squeezing between two qumodes can reach up to $0.82\pm 0.08$ dB. The distribution of the squeezing levels is influenced by the group-velocity dispersion of the microresonator, and notable degradation near qumode 27 is caused by interactions with other transverse modes.

\medskip

\noindent\textbf{2D cluster quantum microcombs}

\noindent We set the primary pump line at mode -1 and employ two secondary pump lines at mode indices \(0\) and \(1\) with powers constituting 20\% and 16\% of the primary pump line, respectively, for generating a 2D lattice cluster state. This configuration results in 3 relatively even major TMSs. The covariance matrix for 56 qumodes, spanning mode indices from \(\pm8\) to \(\pm35\), exhibits negative correlations across three sets of \(\hat{x}\) and \(\hat{p}\) quadratures along the off-diagonal directions, corresponding to qumode pairs with indices summing up to -2, -1, and 0 (Fig.~\ref{figure4}a). For the same reason we illustrated above, correlations induced by minor TMSs and BSM TMSs are negligible relative to those induced by major TMSs. We apply the PPT criterion to 6 overlapping subsets of the cluster states, each comprising 12 qumodes (Fig.~\ref{figure4}b). The PPTs are below unity for all bipartitions, demonstrating the inseparability of the cluster state. The graph structure is a 2D lattice with edge weights of -0.5. We measure these nullifiers (Fig.~\ref{figure4}c), observing squeezing of up to \( 0.81 \pm 0.08  \) dB below the shot noise level. This level of squeezing surpasses the maximum two-mode squeezing achievable \(0.49\pm 0.07\) dB within the system. 

\begin{figure}
\centering
\includegraphics[width=\linewidth]{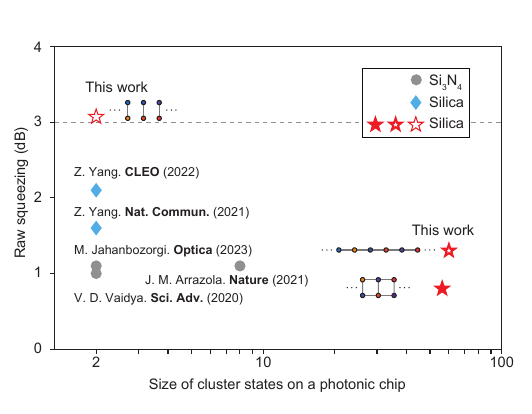}
\caption{{\bf A comparison of the raw squeezing and the size of cluster state in CV quantum states generated on photonic chips.} The dashed line indicates 3 dB value of raw squeezing. Results obtained in Si$_3$N$_4$ microresonators \cite{arrazola2021quantum,vaidya2020broadband,jahanbozorgi2023generation} and silica microresonators \cite{yang2021squeezed,9890809} are plotted for comparison.}
\label{figure5}
\end{figure}

\medskip

{\noindent\bf Discussion}

\noindent Cluster quantum microcombs comprising up to 60 qumodes with various graph structures have been realized. Given that quantum computation necessitates a large scale of cluster states and high levels of raw squeezing \cite{menicucci2006universal,menicucci2014fault}, we compare these performance metrics across several photonic-chip-based CV quantum systems \cite{arrazola2021quantum,yang2021squeezed,jahanbozorgi2023generation,vaidya2020broadband,9890809} (Fig.~\ref{figure5}). The cluster quantum microcombs introduced in this study represent the largest CV cluster states generated on a photonic chip to date. Moreover, the observed 3~dB raw squeezing in the EPR pairs marks a significant advancement toward satisfying the genuine van Loock-Furusawa inseparability criterion for multipartite entangled systems \cite{van2003detecting}. The size and degree of squeezing achieved by the cluster quantum microcombs can also benefit quantum metrology. For instance, the quantum Fisher information \cite{pezze2018quantum,qin2019characterizing} for the 1D, 60-mode cluster state reaches $81.1$, surpassing the standard quantum limit ($60$ for a 60-mode coherent state). We also theoretically show that increasing the size of the cluster quantum microcomb can progressively enhance its metrological capabilities (see Fig.~S17 in Supplementary Information).

Further improvements in our system are feasible. Due to dispersion, microresonator modes may not perfectly align with the qumodes, limiting resonantly-enhanced nonlinear interactions across a broader spectral range. Dispersion engineering, initially developed for broadband classical microcombs \cite{yang2016broadband}, can be adapted to extend the gain bandwidth of parametric processes for larger-scale cluster quantum microcombs. Moreover, advanced photonic integration is expected to leverage the raw squeezing achieved. Current silicon photonic technologies enable the heterogeneous integration of photodetectors with photonic circuitry \cite{yu2020heterogeneous,tasker2021silicon}, eliminating loss during off-chip detection. Integration with other photonic components such as high-speed modulators \cite{wang2018integrated}, microcomb-based LOs \cite{Kippenberg2018,yu2022integrated}, and spectral shapers \cite{wang2015reconfigurable} would facilitate the creation, control, and detection of cluster quantum microcombs on a single photonic chip. We envision that cluster quantum microcombs with sufficient scale and raw squeezing will become the cornerstone for quantum networks with improved robustness, quantum measurements with enhanced precision \cite{aasi2013enhanced}, and quantum computers with explicit quantum advantages \cite{madsen2022quantum,zhong2020quantum}.


\bibliography{scibib}
\bibliographystyle{Science}

\medskip

\noindent {\bf Acknowledgment} This work was supported by National Key R\&D Plan of China (Grant No. 2021ZD0301500), Beijing Natural Science Foundation (Z210004, Z240007), National Natural Science Foundation of China (92150108, 62222515, 12125402, 12174438), and the High-performance Computing Platform of Peking University. Device fabrication in this work is supported by the Micro/nano Fabrication Laboratory of Synergetic Extreme Condition User Facility (SECUF).\\
\\
\noindent{\bf Author contributions} The project was conceived by W.Z., K.L., and Q.-F.Y. Experiments were designed and performed by W.Z., K.L., Y.W., B.W., and Q.-F.Y. Devices were fabricated by X.Z., Y.C., J.L., Z.L., and B.-B.L. Theory was developed by W.Z., Y.W., B.J., F.S., Q.H., and Q.-F.Y. All authors participated in writing the manuscript.\\
\\
\noindent{\bf Competing interests} The authors declare no competing interests.\\
\\
\noindent{\bf Additional information} \\
\noindent{\bf Correspondence and requests for materials} should be addressed to K.L., Q.H., B.-B.L., and Q.-F.Y.\\


\end{document}